\documentstyle[preprint,aps]{revtex}
\begin{document}

\draft

\title{Polaronic optical absorption in electron-doped and hole-doped cuprates}
\author{P. Calvani, M. Capizzi, S. Lupi, P. Maselli, and A. Paolone}
\address{Dipartimento di Fisica, Universit\`a di Roma ``La Sapienza'',
Piazzale A. Moro 2, I-00185 Roma, Italy}
\author{P. Roy}
\address{Laboratoire pour l'Utilization de Rayonnement Electromagn\'etique,
Universit\'e Paris-Sud, 91405 Orsay, France}
\date{\today}
\maketitle
\begin{abstract}
Polaronic features similar to those previously observed in the
photoinduced spectra of cuprates have been detected in the reflectivity
spectra of chemically doped
parent compounds of high-critical-temperature superconductors, both $n$-type
and $p$-type. In Nd$_2$CuO$_{4-y}$ these features, whose intensities depend
both on doping and temperature, include local vibrational modes in the
far infrared and a broad band centered at $\sim$ 1000 cm$^{-1}$.
The latter band is produced by the overtones of two (or three) local modes and
is well described in terms of a
small-polaron model, with a binding energy of about 500 cm$^{-1}$.
Most of the above infrared features are shown to survive in the metallic
phase of Nd$_{2-x}$Ce$_x$Cu0$_{4-y}$, Bi$_2$Sr$_2$CuO$_6$, and
YBa$_2$Cu$_3$O$_{7-y}$, where they appear as extra-Drude peaks.
The occurrence of polarons is attributed to local modes strongly coupled
to carriers, as shown by a comparison with tunneling results.
\end{abstract}

\section{Introduction}
In the long quest for an understanding of the mechanisms of high temperature
superconductivity, the optical properties of cuprates have been extensively
studied, both in the normal and in the superconducting
phase.\cite{Timusk} Electronic transitions which are common to different
families of superconductors have been searched and investigated.
Three well defined structures have been identified in the reflectivity spectra:
The charge-transfer, CT, band (in the visible range), the
MIR band and the $d$ band (both of which in the mid-infrared range).

The CT band is universally attributed to charge-transfer transitions
between O$_{2p}$ and Cu$_{3d}$ states. It has been observed in the insulating
phases of Y-Ba-Cu-O, Bi-Sr-(Ca,Y)-Cu-O, La-Sr-Cu-O, and M-Ce-Cu-O (M = Nd, Pr,
Sm, Eu, Gd). The CT band appears to be associated with excitonic and
polaronic effects.\cite{Falck92}
Upon doping, either by chemical substitution or by photocarrier injection, it
looses spectral weight in favor of a novel band,
the MIR band, which appears inside the CT gap and is normally
absent in stoichiometric samples.\cite{Uchida,Cooper90',Lupi}

A number of experiments have been done, aimed at relating the insurgence
of the MIR band with
superconductivity. In particular, an inverse relation has been
found to hold between the critical temperature and the MIR-band peak
energy,\cite{Foster,CalvaniSSC} which suggests that high-$T_c$
superconductivity is favored by a shift toward lower energy of the MIR band.
This band indeed results from the overlap of two different
components:\cite{Lupi,Thomas92} A contribution independent of temperature
and centered in most compounds around 4000 cm$^{-1}$, another contribution
strongly dependent on temperature and peaked at $\sim$ 1000 cm$^{-1}$.
The latter band, reported for both {\it e-} and {\it h-}doped
systems,\cite{Lupi,Thomas92,Falck93}
is very similar to a structure observed in photoinduced absorption spectra of
different compounds, and already shown to increase in strength with
the critical temperature of the
cuprate.\cite{Epstein-Taliani,Kim,Mihailovic90,Taliani} The
shift observed in the MIR peak as the critical temperature ($T_c$)
increases could then be
explained by the growth of the 1000 cm$^{-1}$ component.
In Nd$_{2-x}$Ce$_x$CuO$_{4-y}$ (NCCO), the strength of this latter increases
at 300 K
with chemical doping and with the concentration of oxygen
vacancies.\cite{Lupi,SSC94} The 1000 cm$^{-1}$ feature has been
therefore attributed to transitions involving defects, whence its
name, $d$ band. Nevertheless, a similar infrared band has been observed
also in {\it h-}doped, oxygen
enriched La$_2$CuO$_{4+y}$ (LCO), where both its lineshape and
temperature dependence are consistent with the absorption from a polaronic
impurity state.\cite{Falck93} The polaronic origin of the $d$ band has been
firmly established in Nd$_2$CuO$_{4-y}$ (NCO). Here the $d$ band has been
resolved
and a relation between its fine structure and some extra-phonon peaks
observed in the far infrared has been found to hold.\cite{SSC94,Cambridge}
Furthermore,
a broad spectrum of experimental techniques, ranging from Extended X-ray
Absorption Fine Structure,\cite{Bianconi} to electron diffraction,\cite{Chen}
and to neutron scattering\cite{Billinge} points toward the existence of polaron
superstructures in superconducting cuprates as well as in non-superconducting
perovskites. It has been also suggested\cite{Bianconi} that such polaron
ordering gives rise to unidimensional conductivity, thus enhancing the critical
temperature of the material.

On the theoretical side, in the early eighties Chakraverty\cite{Chakraverty}
and Alexandrov and Ranninger\cite{Ranninger81} suggested that
polarons would lead to high $T_c$ superconducting phases in ionic
compounds. After the discovery of high-$T_c$ superconductivity, several
authors have again proposed that this phenomenon could be
explained in terms of polaron-related
models,\cite{Emin,Robaszkiewicz,Ranninger} an issue still under discussion.
The influence of polarons on the oxide optical properties has been
analysed in several theoretical works\cite{Reik}
after the early work of Holstein.\cite{Holstein}
Recently, local modes coupled to electrons and holes have
been predicted to show out in the normal-state phonon spectra of high critical
temperature superconductors (HCTS)
for intermediate or large values of the electron-phonon
interaction.\cite{Yonemitsu} The optical conductivity
in the presence of a strong
electron-phonon interaction has been calculated by Emin\cite{Emin} in terms of
the Reik's\cite{Reik} standard polaron model, and by Alexandrov
{\it et al.} by the exact solution of a finite size Holstein
model.\cite{Alexandrov}

This paper is aimed i) at providing clear evidence of polarons in both
hole- and electron-doped cuprates, ii) at showing that they survive in the
metallic
phase, and iii) at looking for hints of their relevance in the superconducting
mechanism. This will be done by
discussing new reflectivity measurements in the far- and mid-infrared
on different cuprates, as well as by a careful reanalysis of
optical and transport results obtained by different authors. In particular, it
will be shown that the additional far-infrared modes observed here in slightly
{\it e-} and {\it h-}doped compounds correspond to the extra-Drude features
observed by several authors in the corresponding superconductors. It will be
also shown that the extra-phonon features determined from the far-infrared
reflectivity in insulating NCO correspond to peaks in the
spectral function $\alpha^2$F($\omega$), as obtained by tunneling
measurements\cite{Huang} in superconducting Nd$_{2-x}$Ce$_x$CuO$_{4-y}$.

\section{Experiment}

Several {\it e-}doped and {\it h-}doped single crystals have been studied, as
listed in Table I. The NCO samples were
prepared by different procedures, in order to obtain crystals with different
concentrations of oxygen vacancies. Sample \#0, prepared at the University
of Geneva, is stoichiometric. Sample \#1, prepared at
AT\&T Bell Labs. of Murray Hill, is as-grown and nearly stoichiometric
($y<0.005$). Samples \#2 and \#3, prepared at the University of Geneva,
were reduced during and after their growth, respectively (see Table I and Ref.
\onlinecite{SSC94}). The samples were mounted on the cold finger of a two-stage
closed-cycle cryostat, whose temperature was kept constant within $\pm$ 2 K and
could be varied from 300 to 20 K. The reflectivity R($\omega$) of all samples,
relative to gold- and aluminum-plated references, has been measured with the
electric field polarized in the $a-b$ plane. Data were collected in two
different laboratories by rapid scanning interferometers, typically from 100
through 25,000 cm$^{-1}$. The real part of the optical conductivity
$\sigma(\omega)$ has been usually obtained from canonical Kramers-Kronig
transformations. A Drude-Lorentz fit has been used to extrapolate the
reflectivity data beyond the measured range. The same fitting procedure
has been used to extract oscillator energies, linewidths, and intensities
from the experimental $\sigma(\omega)$.

\section{Results and discussion}

The reflectivity spectra of the $d$ band, taken at different
temperatures, are reported in Fig. 1 between 100 and 1200 cm$^{-1}$
for three NCO single crystals. The corresponding optical
conductivities, as obtained by a standard Kramers-Kronig analysis, are shown in
Fig. 2.  The $d$ band depends on doping at room temperature, while it is
roughly constant at low $T$. In all samples, and {\it over the whole energy
range}, the reflectivity increases with decreasing temperature, even if the
intensity at high frequency saturates at roughly 200 K, as shown in Fig. 1(b).
In the following subsections we shall separately examine the spectral regions
below 600 cm$^{-1}$ and above 600 cm$^{-1}$ for slightly doped, insulating
cuprates. Then, we shall extend the discussion to their metallic phase.

\subsection{The insulating phase: The $d$ band}

\subsubsection{Description in terms of a standard polaron model}

The data of Fig. 1, already partially reported in Ref. \onlinecite{SSC94},
will be here reanalysed in order to support
the polaronic origin of the band at 1000 cm$^{-1}$ ($d$ band).
Indeed, a polaronic origin is consistent with the temperature dependence
of the linewidth $\Gamma_d$ of this band. According to a conventional adiabatic
approach,\cite{Flynn} a small polaron absorption is described by a single
Gaussian with a linewidth $\Gamma_d$ given by

\begin{equation}
\Gamma_d = \hbar\omega^* [8 ln2 S coth(\hbar\omega^*/2kT)]^{1/2}\;,
\end{equation}

\noindent
where $S$, the Huang-Rhys factor, is a measure of the electron-phonon
interaction, and $\omega^*$ is an average phonon energy. The
optical conductivity $\sigma(\omega)$ of the strongly doped sample \#3 in Fig.
2(c) has been fit at different temperatures by using both Lorentzian
lineshapes (for the far-infrared oscillators up to 700 cm$^{-1}$,
see subsection B), and a single Gaussian (for the broad $d$ band). The fitting
curves for $T$ = 20 K and $T$ = 300 K are shown as dotted lines in Fig. 2 (c).
The resulting values for the $d$-band width $\Gamma_d$ are reported in Fig. 3,
as well as their best fit to Eq. (1) (solid line). One may notice that
Eq. (1) well reproduces  the ``saturation'' observed at low temperature
in the change of the linewidth. The Huang-Rhys factor $S$ turns out to be 4.4
$\pm$ 0.4, indicating that the electron-phonon interaction ranges from medium
to strong. The average phonon energy $\omega^*$ is 210 $\pm$ 10 cm$^{-1}$.
Under the same standard approach, to first order the peak energy
$E_{peak}$ of the polaron band is simply $E_i$ + $S\hbar\omega^*$, where
$E_i$ is the binding
energy of the impurity which provides the free carriers (here, an oxygen
vacancy). In the present case, from the data of Figs. 2 and 4(a)
$E_{peak}$ $\sim$ 800
cm$^{-1}$ in sample \#3, which gives $E_i$ $\simeq$ 0 by using the above
values for $S$ and $\omega^*$. In photoinduced absorption experiments on NCO,
where carriers are not bound to impurities, one
finds\cite{Kim} a band with $E_{peak}$ $\sim$ 1280 cm$^{-1}$,
consistent with the values found, here
and elsewhere,\cite{Thomas92} in samples with low chemical doping.
This indicates that: i) $E_{peak}$ depends on doping, as confirmed in Figs.
2 and 4(a); ii) the impurity binding energy vanishes for both photoinduced
and chemically induced carriers.
As far as the dependence on temperature of the $d$-band intensity is
concerned, the growth of this latter for decreasing $T$ (with a saturation at
$T$ $\approx$ 200 K) is also explained by the conventional adiabatic approach
and is compatible with a small polaron model (the intensity of large polarons
is predicted to be independent on temperature\cite{Emin}).

It is worth noticing that results similar to those presented here are obtained
whenever a polaronic one-phonon model is applied to infrared bands observed at
$\sim$ 1000 cm$^{-1}$. Photoinduced absorption measurements in
YBa$_2$Cu$_3$O$_{6.3}$ and Tl$_2$Ba$_2$Ca$_{0.98}$Gd$_{0.02}$Cu$_2$O$_8$
yielded $\omega^* \simeq$ 200 cm$^{-1}$ in both compounds, with $S$ = 7 and
$S$ = 5.6, respectively.\cite{Mihailovic90} Conversely, reflectivity
measurements in La$_2$CuO$_4$ gave $\omega^* \simeq$ 350 cm$^{-1}$, from
an analysis of the $T$-dependence of the charge-transfer gap in terms
of a Fr\"ohlich polaron,\cite{Falck92} and $S$ = 2.5 from a lineshape fit of
the $d$ band.\cite{Falck93}

The dependence of the $d$ band on doping and temperature is evident in Fig. 2.
Therein, one may observe that the $d$-band intensity at room temperature grows
with doping from (a) to (c), while it does not change appreciably at low $T$.
This is more evident in Fig. 4, where the differences

\[(\Delta\sigma)_T = \sigma(\omega,T,y) - \sigma(\omega,T,0)\]

\noindent
evaluated at $T$ = 300 K for samples \#1, \#2 and \#3 relative to sample \#0,
are plotted. Sample \#0 is the most stoichiometric crystal available ($y$ = 0),
as confirmed by the absence of any sizable $d$-band contribution. Figure 4(a)
shows the insurgence upon doping of a $d$ band, whose peak moves toward low
energy as doping increases from sample \#1 to sample \#3.\cite{nota0,Skantha}
At low doping, the peak is placed above 1200 cm$^{-1}$ at $T$ = 300 K, in
agreement with previous determinations,\cite{Thomas92} while a minimum is
observed at $\sim$ 1000 cm$^{-1}$. This minimum, whose origin is not clear,
is observed only in sample \#1 and rapidly disappears on going to low
temperature. This can be better seen in Fig. 4(b) where the dependence of the
$d$ band on temperature is evidentiated by reporting the normalized difference

\[ (\Delta\sigma/\sigma)_y =  [\sigma(\omega,20,y) - \sigma(\omega,300,y)]/
\sigma(\omega,300,y)\]

\noindent
evaluated at fixed oxygen concentration for all samples. As one can see, the
nearly
stoichiometric sample \#1 shows the highest $(\Delta\sigma/\sigma)_y$, while
for the strongly doped samples this quantity is very small. In turn, samples
\#2 and \#3 show a shift of spectral weigth to lower energies as temperature
decreases, similar to that observed above for increasing doping.

\subsubsection{Departure from a standard polaron model}

In the low doping regime, sample $\#1$ in Figs. 2 and 4, Infrared Active
Vibrations (IRAV) appear on the top of the $d$ band. The energies of the IRAV
are reported in Table II.
The observation of a fine structure with peaks separated by energies of the
order of those of the lattice vibrations  provides further
evidence for the polaronic origin of the $d$ band. Indeed, it has
been recently shown\cite{Alexandrov} that infrared bands exhibiting IRAV like
those in Fig. 4 are consistent with calculations of the polaron optical
conductivity by a finite-size Holstein model, when the conventional adiabatic
approximation is abandoned. In fact, the theoretical $\sigma (\omega)$
exhibits a number of peaks which are separated by phonon-like energies and
have different spectral weights according to the number of phonons involved.

Although we have discussed in detail the {\it e-}doped cuprate family,
the insurgence of a polaronic $d$ band with IRAV is
quite general, as shown in Figs. 5 and 6. Therein,
the mid-infrared spectra of two insulating samples belonging to
different cuprate families are shown for different temperatures.
In Fig. 5, the mid-infrared spectra of an as-grown Gd$_2$CuO$_4$ (GCO) single
crystal are reported for different temperatures. This sample also exhibits
a well defined $d$ band extending up to
$\approx$ 2000 cm$^{-1}$, whose intensity strongly increases for decreasing
temperature and nearly saturates at 200 K, as in NCO. The energies of the
IRAV on the top of the $d$ band are reported in Table II.

The mid-infrared reflectivity of a single crystal of insulating
Bi$_2$Sr$_2$YCu$_2$O$_8$ (BSYCO) is shown in Fig. 6 for three different
temperatures. The spectra in Fig. 6 are affected by a
large uncertainty in the absolute intensity values, due to the small dimensions
of the sample. However, together with a phonon peak\cite{Pal} at 637
cm$^{-1}$, one may observe a well defined $d$ band with a few peaks on the top.
Their energies are reported in Table II.

In conclusion, $d$ bands with IRAV on their top are observed
in both Gd$_2$CuO$_{4-y}$ and Bi$_2$Sr$_2$YCu$_2$O$_8$,
at energies corresponding to those found in Nd$_2$CuO$_{4-y}$. This
supports a common origin for polarons in these materials, whose composition
is completely different except for having
Cu-O planes. In the following subsection, a relation will be established
between the energies of the peaks resolved in the $d$ band, and those
of the extra-phonon modes induced by doping and detected in the same spectra
at lower energies.

\subsection{The insulating phase: Local vibrational modes}

The far-infrared, room-temperature $\sigma(\omega)$ of the NCO stoichiometric
sample \#0, already published in a previous paper,\cite{Lupi} is reported in
Fig. 7(a) for the reader's convenience. It represents a good example of a
``purely phononic'' far-infrared spectrum, where only
the four infrared-active E$_u$ phonons predicted by group theory for
the $T'$ structure of NCO are detected.
Their energies are reported in the bottom part of Table II.
The far-infrared $\sigma(\omega)$ at 20 K is presented in Fig. 7(b) and 7(c)
for the NCO crystals \#2 and \#3, respectively. In Fig. 7(d),
$\sigma(\omega)$ is instead reported at 10 K for an NCO sample annealed under
Ar atmosphere and measured by different authors.\cite{Heyen} The plot has been
obtained by multiplying by $\omega/(4\pi)$ the imaginary part of the
dielectric function $\epsilon_2$, as reported in Fig. 1 of
Ref.\onlinecite{Heyen}.

By a first inspection of Fig. 7, one sees that the four E$_u$ phonon peaks of
Fig. 7(a) evolve into four bands when passing to the low temperature spectra
of the doped samples (Figs. 7(b) to 7(d)). Those bands are structured and in no
way can be reduced to single-phonon modes, even if they are centered roughly
at the energies of the four E$_u$ vibrations.

In addition, one observes at low $T$ a broad background between 180 and 250
cm$^{-1}$. Its intensity increases as temperature is lowered but decreases
for increasing doping. The latter feature shows that the origin of this
background, which will not be discussed further
on, is not related to excess charges.

A weak peak at 165 cm$^{-1}$ emerges barely from the background in Fig.
7(b), while it is more clear in Fig. 7(d) (see asterisk) as well as in the
spectrum at 10 K of a NCO single crystal reported in Fig. 3 of Ref.
\onlinecite{Crawford}. It has been attributed to magnetic effects
in Ref.\onlinecite{Heyen}, to disorder-induced effects in
Ref.\onlinecite{Crawford}. It should be mentioned that neutron scattering
data show a transition at 165 cm$^{-1}$, which has been attributed to
a crystal field splitting\cite{Pyka} in the electronic levels of the
Nd$^{3+}$ ion.  However, this transition is not
expected to be infrared active and it has not been
observed\cite{Thomas93} in stoichiometric samples down to $T$ = 10 K. This
rules
out the hypothesis of a magnetic origin .

An explanation for the 165 cm$^{-1}$ peak should then be searched in the
context of the profound modifications induced by doping on the phonon spectrum.
Indeed, at low $T$ a number of IRAV add to the E$_u$
phonon peaks in the reduced crystals \#2 and \#3, as well as in the
far-infrared spectra at low $T$ of  Refs. \onlinecite{Heyen} and
\onlinecite{Crawford} (not discussed by the authors).
These IRAV modes are similar to the Photoinduced Local Modes (PILM) observed by
optical injection of
charges.\cite{Epstein-Taliani,Kim,Mihailovic90,Taliani,Mihailovic91}
This rules out the possibility that the IRAV observed
in the present, chemically doped, crystals might be simply due to point
defects,
e.g., oxygen vacancies.

The attribution of the IRAV bands in Fig. 7 to local
modes\cite{SSC94} is supported by the following arguments.

The energies of IRAV show just a few accidental correspondences with
the infrared longitudinal modes of the stoichiometric compound,\cite{Lupi}
or with the Raman-active modes\cite{Heyen} of the $a-b$ plane in the same
material. Moreover, the high orientation of the
present single crystals excludes the possibility of direct contributions
from $c$-axis vibrations.  Fano antiresonances between some
$c$-axis phonons and the electronic continuum  of the $a-b$
plane\cite{Timusk90} can also be excluded, due to the absence
of any sizable Drude contribution in the present NCO samples at low $T$.
On the other hand, the IRAV of Fig. 7 can be
distinguished from the four E$_u$ phonons of the $a-b$ plane by their increase
with doping\cite{SSC94,Thomas93} (in NCO with the oxygen deficiency $y$)
and their decrease with temperature between 200 and 300 K (at $T$ $<$ 200 K the
intensity of most extra modes remain constant).\cite{SSC94} This suggests
that the IRAV may arise from lattice
distortions induced in the polar lattice of NCO by the excess charges created
by doping.
Recent calculations in Cu-O clusters doped by holes show that the total energy
is minimized when the hole resides on a Cu atom and the four nearest neighbour
O atoms move toward the Cu atom.\cite{Yonemitsu} This doping induced lattice
distortion gives rise to new infrared active (local) modes. They will appear as
satellite bands on both the low- and the high-energy sides of the unperturbed
lattice phonons, with intensities which increase with the strength of the
hole-phonon interaction.\cite{Yonemitsu}

On the grounds of the above considerations, in the strongly doped samples
the optical conductivity $\sigma(\omega)$ can be split into three
contributions

\[ \sigma(\omega) = \sigma_{ph}(\omega) + \sigma_{BGD}(\omega) +
\sigma_{LM}(\omega)\]

\noindent
where $\sigma_{ph}(\omega)$ is the phonon term and
$\sigma_{LM}(\omega)$ is that of the local modes (LM). $\sigma_{BGD}(\omega)$
is the contribution from the broad background around 200 cm$^{-1}$. As already
stated, this latter depends on doping in the opposite way to that found
for the local modes. The oscillator parameters have been obtained by
a Drude-Lorentz fitting procedure.\cite{Lupi} The oscillator energies,
listed in Table II, result from an average over the
Nd$_2$CuO$_{4-y}$ samples \#2 and \#3. Most of these values can be
interpreted in terms of overtones or combination bands of two fundamental
modes, as explained in the following.

The local-mode conductivity $\sigma_{LM}(\omega)$ at 20 K is reported in
Fig. 8 for sample \#2.
Several bands are well resolved, whose peak energies
have been partially reported previously.\cite{SSC94,Grenoble}
Surprisingly, in both samples \#2 and \#3, the total intensity of LM is
roughly twice as much as the total intensity of the four E$_u$ phonons.
It has already been mentioned that these IRAV show a dependence
on doping and temperature similar to that found for the $d$ band.
It is therefore most likely that these modes are the low order elements of a
polaron series whose convolution gives rise to the $d$ band.
The structures still observed on the top of this band at intermediate
doping, see \#2, would then be the remnants of overtone and combination bands
of a few of these far-infrared LM. Higher-energy overtones would be too broad
to be resolved.

The two lowest-energy LM provide the basis for a simple model which i) accounts
for most of the far-infrared and the $d$-band IRAV,
ii) confirms the interpretation of the spectrum in terms of
small polarons, and iii) provides a few characteristic parameters of the
polaron. The two lowest LM energies,\cite{Nota}
$\omega_1$ = 165 cm$^{-1}$ and $\omega_2$ = 287 cm$^{-1}$, lead to a mean
value $\omega^* = (I_1\omega_1+I_2\omega_2)/(I_1+I_2)$ = 260 $\pm$ 10
cm$^{-1}$ at 20 K in \#2. This value may be compared with that (210 $\pm$ 10
cm$^{-1}$) obtained above  by fitting to Eq. (1) the $d$-band linewidth
of sample \#3 derived in a single-phonon approach.

Within the adiabatic model, the (thermal) binding energy E$_p$ of the small
polaron, as described by a standard single-phonon calculation, can be
estimated. The optical
conductivity is then given by the Reik's formula\cite{Reik}

\begin{equation}
\sigma (\omega) \propto [t^2/(2E_pT)][(1-e^{-\omega /T})/\omega]
exp[-(\omega-2E_p)^2/(8E_pT)]\:,
\end{equation}

\noindent
where $t$ is an intersite hopping frequency. According to Eq. (2), for
$T \to 0$ the maximum absorption will occur at
$\omega \simeq 2E_p$, namely at $\omega \simeq n \omega^*$
($n>1$ is the number of phonons of energy $\omega^*$ which best approximates
the Huang-Rhys factor S). From an inspection of Fig. 4(a),
where the $d$-band peak moves to low energy for increasing doping,
$E_p$ turns out to be $>$ 600 cm$^{-1}$ in sample \#1, $\sim$ 550 cm$^{-1}$ in
sample \#2, and $\sim$ 400 cm$^{-1}$ in the most heavily doped sample \#3.
Alternatively, with a phonon energy $\omega^*$ = 260 $\pm$ 10 cm$^{-1}$,
Eq. (2) provides an
estimate of the polaron binding energy $E_p$ from the lineshape of the
$d$ band (see Fig. 5 in Ref.\onlinecite{Emin}). Referring to Fig. 4 of the
present work, one obtains $E_p \leq$ 3 $\omega^*$ $\simeq$ 780 cm$^{-1}$ for
sample \#2, $E_p \sim$ 2 $\omega^*$ $\simeq$ 520 cm$^{-1}$ for sample \#3,
in fair agreement with the previous determinations. The decrease of the polaron
binding energy for increasing doping is consistent with the temperature
dependence of the $d$ band in Fig. 4(b), which becomes smoother as
going from sample \#1 to samples \#2 and \#3.
Finally the saturation temperature for the $d$ band intensity for NCO
($\approx$ 200 K) is also well accounted for by the above polaron parameters.

The energies $E_j$ of almost all the peaks in Fig. 8 are found to be given
either by $n_1$$\omega_1$ or by $n_2$$\omega_2$, for $n_1$ and $n_2$ as listed
in Table II. As expected, the energies of most overtone bands are shifted with
respect to their algebraic values by negative amounts due to anharmonic
effects which increase with the overtone energy. If the
overtones and combination bands of the two fundamentals, with linewidths
$\gamma_j$ given by Eq. (3) (see below), are used, the entire band
can be described up to 2000 cm$^{-1}$. As shown in Table II, a few of the
observed IRAV cannot be
explained in terms of overtones of the two lowest local modes 1 and 2. These
peaks can either be due to combination bands of 1 and 2, as suggested
previously,\cite{SSC94} or they can be assigned to a third polaron series
originating from a mode at $\omega_3$ = 375 cm$^{-1}$, as indicated in the
same Table.

In Figure 9, the linewidths $\gamma_j$ of the $j$-th overtone in
the $d$ band entering the fit of the far- and mid-infrared spectra of NCO
\#2 are plotted as a function of the oscillator peak energies $\omega_{pj}$.
Except for peak 2 in Fig. 8, whose anomalous broadening is possibly related to
the proximity of the broad background mentioned above, the linewidths follow
the phenomenological relation (solid line in the Figure)

\begin{equation}
\gamma_j = 10 + 1.6 \times 10^{-7} \omega_{pj}^{3} \;,
\end{equation}

\noindent
where $\gamma_j$ and
$\omega_{pj}$ are given in cm$^{-1}$ (a smoother dependence on $\omega_{pj}$
had been reported in Ref. \onlinecite{SSC94}, on the basis of preliminary
results). As suggested by a direct inspection of data for sample \#2 in Fig. 9,
a discontinuity appears at $\sim$ 600 cm$^{-1}$. At lower energies,
the oscillator bands are narrow, with almost constant linewidths, while at
high energies they rapidly broaden. This behavior can be explained
on the basis of the small polaron model, by assuming that the IRAV
at low energies are due to the excess electrons being localized
and vibrating in a single potential well. The high-order IRAV
forming the $d$ band for $\omega_{pj} >$ 600 cm$^{-1}$ would
instead promote intersite jumps of the excess charge. This situation
corresponds\cite{Alexandrov} to a small polaron hopping at a frequency
$t \approx E_p$ $\approx$ 600 cm$^{-1}$. This value is consistent with
the independent evaluations of
$E_p$ presented above for sample \#2 ($>$ 600 cm$^{-1}$ and $\sim$ 550
cm$^{-1}$). On the other hand, the small values of the low-energy overtone
linewidths imply that the dispersion of the fundamental modes is small,
consistently with their attribution to local vibrations. In this respect, the
IRAV here identified do not represent a single case.
Several extra-phonon modes were observed by neutron scattering in
YBa$_2$Cu$_3$O$_{7-y}$ (YBCO).\cite{Arai} The one at 700 cm$^{-1}$ is
particularly strong and has been
studied in detail. It did not show any appreciable wavevector dependence
and therefore has been attributed to a local vibration.

As far as the intensities of the IRAV are concerned, they provide
further evidence for the polaronic origin of the $d$ band. Indeed, the
intensities of the overtones are comparable with, or even higher than,
the intensities of the fundamental modes 1 and 2. As a matter of fact,
the strength of vibrational overtones resulting from transitions within an
anharmonic potential normally decreases by orders of magnitude as the order
of the overtone increases. This is not the case for strong electron-phonon
coupling. On the other hand, the LM are produced by the introduction of
a few percent of oxygen vacancies into the lattice. One can wonder how
the intensities of the doping-induced local modes may be comparable with
those of the extended $E_u$ modes of the lattice. However,
the charged distortion created in the lattice by the excess carriers may
induce huge local dipole moments which in turn lead to intense infrared
absorption. This has been indeed predicted for the Cu-O
plane in the case of strong electron-phonon coupling.\cite{Yonemitsu}

The present results for the far-infrared spectrum of the {\it e-}doped
insulators are confirmed by the $\sigma(\omega)$ reported for the
{\it h-}doped La$_2$CuO$_{4+y}$ in Ref. \onlinecite{Bazhenov} and in
Ref. \onlinecite{Thomas93}. The latter data are shown in Fig. 10.
One may notice that the onset of the $d$ band is found at
$\sim$ 600 cm$^{-1}$, as in NCO, in spite of the large difference in the
phonon energies of these two materials. At lower energies, a peak at about
500 cm$^{-1}$ in LCO is the counterpart of that at 480 cm$^{-1}$ in NCO, see
Fig. 7. Both these structures strongly increase for decreasing temperature
and for increasing doping. The same holds for the extra-phonon peaks
around 250 cm$^{-1}$.\cite{Thomas93,Bazhenov}
Any closer comparison between NCO and LCO in the far infrared is
made difficult by different contributions from the background in the two
systems.

\subsection{The metallic phase}

It has been shown in the previous Sections that IRAV and polaronic bands
are observed in the far- and the mid-infrared spectra, respectively,
of the insulating parent compounds of HCTS. In the present Section we extend
our analysis to the metallic phase, by showing
preliminary results on a {\it hole-}doped compound and by discussing
spectra reported in the literature for both {\it electron-}doped and
{\it hole-}doped cuprates. Those spectra exhibit unexpected features
that are superimposed to the free-carrier absorption (Drude term). We show
that in most cases these features correspond to the IRAV peaks
and to the $d$ band observed in the insulating cuprates. We conclude that
polaron bands are likely to survive in several metallic cuprates.

Evidence for extra-Drude features in superconducting
Bi$_2$Sr$_2$CuO$_6$ (BSCO) is provided in Fig. 11. The reflectivity at room
temperature of a superconducting 2-$\mu$m-thick Bi$_2$Sr$_2$CuO$_6$ film
($T_c$ = 20 K) is reported in the top part of the
Figure. The film (\#6) was grown by liquid phase epitaxy on LaGaO$_3$.
The optical conductivity of the same sample, as obtained by Kramers-Kronig
transformations of R($\omega$), is reported in the bottom part of the Figure.
As shown in Table III, the broad features superimposed to the Drude term
in the far infrared have energies which correspond to those of the IRAV
found in NCO. It should be noticed that the room-temperature Drude
term in the spectrum of Fig. 11 is comparable with, or weaker than, the local
mode contributions. Indeed, IRAV should be detected more easily in systems
with low carrier densities, like for instance that of Fig. 11. The existence
of extra-Drude features in this system is confirmed by the
mid-infrared reflectivity spectrum of a BSCO single crystal
with $T_c$ = 10 K, which exhibits a structured $d$ band.\cite{Grenoble}

The $d$ band at $\sim$ 1000 cm$^{-1}$ has
been already observed in several metallic and superconducting cuprates.
Indeed, it has been shown\cite{Lupi} that the infrared optical conductivity
of metallic NCCO can be decomposed into three contributions:
i) a normal Drude term; ii) a $d$ band peaked at $\approx$ 1000 cm$^{-1}$;
iii) a MIR band at $\approx$ 4000 cm$^{-1}$.
In a similar way, in four YBCO single crystals with different oxygen contents
and critical temperatures $T_c$ ranging from 30 to 90 K, Thomas
{\it et al.}\cite{Thomas91} have decomposed  $\sigma(\omega)$ into:
i) a normal Drude term; ii) a low-energy, $T$-dependent contribution, peaked at
$\approx$ 1000 cm$^{-1}$, which can be straightforwardly identified with the
present $d$ band; iii) a $T$-independent MIR band at $\approx$ 5000 cm$^{-1}$.
Moreover, far-infrared structures clearly appear in the low temperature optical
conductivity of the samples with $T_c$ = 50 K and $T_c$ = 80 K $\relbar$
see Fig. 8 in Ref. \onlinecite{Orenstein}. Their energies are close to those
reported here for the IRAV of insulating NCO, as shown in Table III for
the sample with $T_c$ = 80 K.

Extra-Drude features can be identified also in the infrared spectra
of a Nd$_{1.85}$Ce$_{0.15}$Cu0$_{4-y}$ thin film\cite{Hughes} with
$T_c$ = 21 K, and of a {\it h-}doped, metallic\cite{Kamaras}
YBCO, with $T_c$= 90 K.
A broad extra-Drude contribution, reminescent of the
$d$ band, is present in both compounds with an onset at $\sim$ 600 cm$^{-1}$.
IRAV modes are reported in the far-infrared, whose energies do
not correspond to those expected for phonons in such
materials. Once again, most of those IRAV energies correspond to the
NCO extra modes, see Table III. Finally, in the same Table we report for
comparison all the
modes observed by neutron scattering in YBCO\cite{Pyka,Arai} which
coincide neither with the Raman, nor with the infrared phonons.

One may remark that the mode reported to occur at $\sim$ 245 cm$^{-1}$ in the
infrared spectra of BSCO, NCCO, and YBCO
is missing in the NCO spectra. As already mentioned, however, the feature
at 282 $\pm$ 4 cm$^{-1}$ in NCO exhibits an unusually large full width at half
maximum (40 cm$^{-1}$, see Fig. 9), which could be due to the superposition of
two distinct modes.

Finally, it may be worth to recall here the results of tunneling experiments
on a Nd$_{1.85}$Ce$_{0.15}$CuO$_{4-y}$ sample with $T_c$ = 22 K.
Huang {\it et al.}\cite{Huang} have measured at 4.2 K the spectral function
$\alpha^2 F(\omega)$ in two different tunnel junctions, finding maxima at the
energies reported in Table III.
One can easily see that those maxima coincide in energy with the
infrared extra-phonon peaks found in NCO.
Only accidental correspondences can be found, instead, between
$\alpha^2 F(\omega)$ and the infrared- or the Raman-active phonons.
The feature at 125 $\pm$ 6 cm$^{-1}$, which is not included among the
IRAV of NCO, corresponds to the lowest-energy phonon in the $a-b$ plane of this
cuprate.

\section{Conclusion}

In the present work, evidence for the existence of polaronic
infrared bands in both hole- and electron-doped insulating cuprates has
been provided.
These bands are peaked at $\approx$ 1000 cm$^{-1}$, as those observed in the
photoinduced spectra of several parent compounds of HCTS. Within standard
one-phonon, adiabatic polaron models, they correspond to small polarons
with a binding energy of $\sim$ 500 cm$^{-1}$.  A polaron fine structure
has been partially resolved in the present spectra, similar to that
predicted by recent exact calculations on small clusters. The fine structure
is here explained in terms of overtones of the additional modes, depending
both on doping and temperature, which appear at 165 and 282 cm$^{-1}$ (and
possibly at $\sim$ 375 cm$^{-1}$) in the
phonon spectrum of Nd$_2$CuO$_{4-y}$. These modes are assigned to strongly
infrared-active vibrations of the lattice, locally perturbed by the excess
charges. They also appear in several metallic cuprates, both {\it hole-} and
{\it electron-} doped, where they are
superimposed to the Drude term. In the HCTS several authors have also observed
broad bands, dependent on temperature and centered at $\approx$ 1000 cm$^{-1}$.
These results confirm the link between the local modes in the far infrared
and the $d$ band. They also suggest that the polaronic behavior of the optical
conductivity is a general feature of the Cu-O plane, and indicate
that polarons may survive in the metallic phases of HCTS. Therein,
two types of carriers may be present, small polarons with hopping
energies of the same order as the binding energy ($\approx$ 500 cm$^{-1}$)
and carriers with lower effective masses. Tentatively, the
latter carriers may represent either a normal Fermi liquid coexisting with a
substrate of small polarons, or the coherent part of a polaron fluid.

A crucial question obviously concerns the role polarons may
play in High-$T_c$ superconductivity. On the basis of the present results
and of our interpretation of data already available in the literature, we can
contribute to the debate by proposing the following two considerations.

In a previous Section we have discussed recent tunneling results, which
identify the lattice modes that in NCCO strongly interact with the carriers
responsible for superconductivity. These modes have the same energies
as the IRAV here detected in the insulating phase. This comparison, which is
meaningful as local modes have negligible dispersion in the space of
momenta, suggests that in NCCO {\it the modes which interact most strongly
with carriers in the superconducting phase are those of local origin,
not the extended phonons of the unperturbed lattice}.

Secondly, one should carefully examine the infrared data taken by Thomas
{\it et al.}\cite{Thomas91} in YBCO samples with different $T_c$.
In Ref. \onlinecite{Thomas91}, it has been found
that $T_c$ increases with the effective number of carriers in the $d$ band
$n_d = {2m^*V\over \pi e^2} \int _0^\infty \sigma_d (\omega)\, d \omega$.
The intensity of the MIR band was found instead to be {\it independent of $T_c$
within errors}. Several
authors\cite{Ranninger,Alexandrov} have suggested that a Bose condensation of
bipolarons may account for superconductivity in HCTS. In this case $T_c
\propto n_{bipolaron}^{2/3}$. If one assumes $n_d \propto n_{bipolaron}$,
one could wonder whether $T_c \propto n_d^{2/3}$. Indeed, from the data of
Ref. \onlinecite{Thomas91} one obtains with a good approximation
$T_c \propto (n_d/n_{MIR})^{2/3}$. At present however, we do not know
whether this phenomenological argument is physically meaningful or not.
In particular, one should justify the assumption that the number of effective
carriers in the small-polaron $d$-band is proportional to that of mobile
bipolarons which would condense below $T_c$.

\acknowledgments
We wish to thank A. Bianconi, R. Caciuffo, C. Castellani, B. Chakraverty,
C. Di Castro, M. Grilli, V. Kabanov, and G. Strinati for useful discussions.
G. Balestrino, H. Berger, S-W. Cheong, W. Sadowski, and E. Walker are
gratefully acknowledged for providing the
samples used in this work. We are also indebted to P. G. Medaglia for his
collaboration. Some authors (P. C., A. P., and S. L.) wish to thank
the LURE laboratory, where some of the present data have been collected.

This work has been supported in part by Istituto Nazionale di Fisica della
Materia of Italy and by
{\it Progetto Finalizzato Superconduttivit\`a} of Consiglio Nazionale delle
Ricerche of Italy.



\begin{table}

\caption
{Listing of the samples measured in the present experiment. They are all
single crystals, except for sample \#6 which is a highly oriented film,
2 $\mu$m thick, grown by liquid phase epitaxy.}
\label{Table I}

\begin{tabular}{cccccc}

     Item     &         Material           &     Code       &    Doping $y$   &
    Source                 &               Comments           \\
\tableline
     \#0      &      Nd$_2$CuO$_{4-y}$     &     MN8        &     $\sim$0
&
Univ. of Geneva            &       as grown                   \\
     \#1      &      Nd$_2$CuO$_{4-y}$     &     MN24       &     $<$0.005
&
AT\&T Bell Labs, M. H.     &       as grown                   \\
     \#2      &      Nd$_2$CuO$_{4-y}$     &     MN22       &       $>$0
&
Univ. of Geneva            &      grown in reducing atm.      \\
     \#3      &      Nd$_2$CuO$_{4-y}$     &     MN19       &       0.04
&
Univ. of Geneva            & annealed in N$_2$, 900 C, 10-15 hrs\\
     \#4      &      Gd$_2$CuO$_{4-y}$     &     MG01        &     $<$0.005
&
AT\&T Bell Labs, M. H.     &     as-grown                     \\
     \#5      &Bi$_2$Sr$_2$YCu$_2$O$_{8+y}$&     MBY1        &
&
Univ. of Lausanne          &                                  \\
     \#6      &     Bi$_2$Sr$_2$CuO$_6$    &     FBP4        &
&
Univ. of Rome II           &     as-grown                     \\
     \#7      &     Bi$_2$Sr$_2$CuO$_6$    &     M4B         &
&
Univ. of Lausanne          &     as-grown                     \\
\end{tabular}

\end{table}


\begin{table}

\caption
{Peak energies, in cm$^{-1}$, of the spectral features observed in the
reflectivity spectra of a variety of insulating slightly doped cuprates.
In the case of NCO, the peak energies of the local modes have been obtained by
a best fit of optical conductivity data. For NCO, a tentative
assignement of the local modes is also given in terms of three series of
overtones, as explained in the text.}
\label{Table II}

\begin{tabular}{cccccccc}

NCO \#0 &NCO \#1 &NCO \#2, \#3 & $n_1$  & $n_2$  & $n_3$ &GCO \#4 & BSYCO \#5\\
\tableline
Local mode&      &             &        &        &       &        &          \\
\tableline
        &        &   165       &    1   &        &       &        &          \\
        &        &   282$\pm$4 &        &    1   &       &        &          \\
        &        &   327       &    2   &        &       &        &          \\
        &        &   366       &        &        &   1   &        &          \\
        &        &   383       &        &        &   1   &        &          \\
        &        &   471       &    3   &        &       &        &          \\
        &        &   507       &        &        &       &        &          \\
        &        &   522       &        &        &       &        &          \\
        &        &   536       &        &    2   &       &        &          \\
        &        &   550       &        &    2   &       &        &          \\
        &  630   &   641       &    4   &        &       &   635  &          \\
        &  710   &   717$\pm$7 &        &        &   2   &   715  &    715   \\
        &  780   &   782       &    5   &    3   &       &   800  &    790   \\
        &  855   &   861$\pm$2 &        &        &       &   870  &    865   \\
        &  895   &             &        &        &       &        &          \\
        &        &   960$\pm$3 &    6   &        &       &        &          \\
        & 1020   &             &        &    4   &       &        &          \\
\tableline
Phonon Mode&     &             &        &        &       &        &          \\
\tableline
  132   &        &   132       &        &        &       &        &          \\
  309   &        &   305       &        &        &       &        &          \\
  352   &        &   354       &        &        &       &        &          \\
  511   &        &   514       &        &        &       &        &          \\

\end{tabular}

\end{table}


\begin{table}

\caption
{The peak energies, in cm$^{-1}$, of the spectral features observed in a
variety of superconducting cuprates are compared with those obtained for the
local modes of insulating NCO. Data are extracted from optical conductivities,
tunneling experiments, and neutron scattering experiments.}
\label{Table III}

\begin{tabular}{ccccccc}

NCO \#2, \#3 & BSCO \#6&NCCO$^a$&NCCO$^b$&YBCO$^c$&YBCO$^d$&YBCO$^e$ \\
\tableline
$T_c$ = 0 K    &= 20 K&= 21 K & = 22 K& = 90 K& = 80 K& = 93 K\\
\tableline
Local mode     &      &       &       &       &       &       \\
\tableline
     165       &  180 &       &  165  &  165  &  185  &168$^f$\\
               &      &       &       &  210  &  215  &       \\
               &  245 &  245  &       &       &  240  &       \\
     282$\pm$4 &  270 &  285  &  270  &  280  &       &       \\
     327       &      &  325  &  325  &  345  &  340  &       \\
     366       &  365 &  370  &  375  &       &       &       \\
     383       &  385 &       &       &       &       &  385  \\
               &      &       &  415  &  405  &  405  &       \\
     471       &  485 &       &  470  &  480  &  470  &  460  \\
     507       &      &  490  &       &       &       &  495  \\
     522       &      &       &       &       &       &       \\
     536       &  530 &       &       &       &       &       \\
     550       &      &       &       &  555  &  555  &  550  \\
               &      &       &       &       &       &  595  \\
     641       &  640 &       &       &       &  625  &  630  \\
     717$\pm$7 &      &       &       &       &       &  700  \\
\tableline
Phonon Mode    &      &       &       &       &       &       \\
\tableline
      132      &      &       &  125  &       &  125  &       \\
\end{tabular}

\noindent
$^a$ From $\sigma(\omega)$, Ref. 45.

\noindent
$^b$ From tunneling data, averaged over two different junctions, as reported in
Ref. 28.

\noindent
$^c$ From $\sigma(\omega)$, averaged over two different temperatures, Ref. 46.

\noindent
$^d$ From $\sigma(\omega)$, averaged over different temperatures, Ref. 44.

\noindent
$^e$ From neutron inelastic scattering, Ref. 41.

\noindent
$^f$ From neutron inelastic scattering, $T_c$ = 92 K, Ref. 47.
\end{table}


\begin{figure}
\caption{The reflectivity $R(\omega)$ of three Nd$_2$CuO$_{4-y}$ single
crystals (\#1 to \#3) in the far- and  mid-infrared region, at different
temperatures. The oxygen deficiency $y$ increases from top to bottom.}
\label{fig1}
\end{figure}

\begin{figure}
\caption{The optical conductivities  corresponding to the reflectivity data of
the three NCO samples reported in Fig. 1 (thin solid lines correspond to $T$ =
300 K, thick solid lines to $T$ = 20 K). The dotted lines in (c) are the result
of a best fitting procedure in terms of Lorentz oscillators for the extended
and local modes, and of a Gaussian lineshape for the $d$ band.}
\label{fig2}
\end{figure}

\begin{figure}
\caption{The width $\Gamma_d$ of the $d$ band is plotted for the most doped
Nd$_2$CuO$_{4-y}$ sample \#3 as a function of temperature. The solid line is
the best fit of Eq. (1) to data. The average phonon energy
$\omega^*$ and the Huang-Rhys factor $S$ are given.}
\label{fig3}
\end{figure}

\begin{figure}
\caption{ (a) The difference $(\Delta\sigma)_T$ = $\sigma(\omega,300,y)$ -
$\sigma(\omega,300,0)$ is plotted for the reduced samples \#1, \#2, and
\#3 ($y>0$) relative to the as-grown sample \#0 ($y=0$), in order to subtract
the high energy contributions.
(b) $(\Delta\sigma/\sigma)_y$ =
[$\sigma(\omega,20,y)$ - $\sigma(\omega,300,y)$]/$\sigma(\omega,300,y)$
is reported for three Nd$_2$CuO$_{4-y}$ samples, in order to show the
different
behaviors with temperature of the $d$ band in crystals with different doping.}
\label{fig4}
\end{figure}

\begin{figure}
\caption{The mid-infrared reflectivity $R(\omega)$ for an as-grown
Gd$_2$CuO$_4$ single crystal at three different temperatures. Low-$T$
spectra show a $d$ band partially resolved into peaks.
The features indicated by arrows are instrumental.}
\label{fig5}
\end{figure}

\begin{figure}
\caption{The mid-infrared reflectivity $R(\omega)$ for of an as-grown
Bi$_2$Sr$_2$YCu$_2$O$_8$ single crystal, at three different temperatures.
As in Fig. 4, low-$T$ spectra show in the $d$ band several peaks. The small
dips indicated by arrows are instrumental.}
\label{fig6}
\end{figure}

\begin{figure}
\caption{The real part $\sigma(\omega)$ of the optical conductivity
of four Nd$_2$CuO$_{4-y}$ samples. The spectrum in (d) results
from a re-elaboration of data reported in Ref. [33].}
\label{fig7}
\end{figure}

\begin{figure}
\caption{$\sigma_{LM}(\omega)$, obtained for the doping-dependent additional
modes only, is plotted for sample \#2 at 20 K. The fundamental modes 1
(165 cm$^{-1}$) and 2 (282 cm$^{-1}$) are indicated.}
\label{fig8}
\end{figure}

\begin{figure}
\caption{Linewidth $\gamma$ as a function of the peak energies for both local
modes and spectral structures identified in the $d$ band of NCO. The solid
line is a best fit to data as explained in the text.}
\label{fig9}
\end{figure}

\begin{figure}
\caption{$\sigma(\omega)$ of an oxydized LCO sample, as reported in Ref. [36].}
\label{fig10}
\end{figure}

\begin{figure}
\caption{Top: The room temperature reflectivity $R(\omega)$, in the far- and
mid-infrared region, of a metallic 2-$\mu$m-thick Bi$_2$Sr$_2$CuO$_6$ film
(\#6), with $T_c$ = 20 K. Bottom: The
optical conductivity of the same sample, as obtained by Kramers-Kronig
transformations.}
\label{fig11}
\end{figure}


\begin{references}
\bibitem{nota0}
   It should be pointed out that
   Fig. 4(a) rules out the possibility that the $d$ band might have a
   purely magnetic origin, as one could infer from the observation that its
   peak energy is close to that of a magnon propagating in these materials.
   Indeed, the N\'eel temperature in NCO is lower than 270 K for any doping
   level\cite{Skantha} and is known to decrease rapidly
   in strongly reduced crystals. Therefore, in no way one could explain the
   existence of a well defined magnonic absorption at 300 K, which
   moreover increases with oxygen reduction.
\bibitem{Nota}
   A higher value for $\omega_1$ (183 cm$^{-1}$) has been determined
   in Ref. \onlinecite{SSC94}, where the background at 200 cm$^{-1}$ has
   not been subtracted, although this contribution can be distinguished in the
   reflectivity data of Fig. 1 of Ref. \onlinecite{SSC94}.
\end{references}
\end{document}